\documentclass[twocolumn,amsmath,amssymb]{revtex4}
\usepackage[dvips]{graphicx}
\usepackage{dcolumn}
\usepackage{bm}

\begin{document}
\title[]
{On the feasibility of charged wormholes}
\author{Peter K.\,F. Kuhfittig}
\email{kuhfitti@msoe.edu}
\address{Department of Mathematics\\
Milwaukee School of Engineering\\
Milwaukee, Wisconsin 53202-3109 USA}

\begin{abstract}\noindent
While wormhole spacetimes are predictions of the general theory of
relativity, specific solutions may not be compatible with quantum
field theory.  This paper modifies the charged wormhole model of Kim and
Lee with the aim of satisfying an extended version of a quantum inequality
due to Ford and Roman.  The modified metric may be viewed as a solution of
the Einstein fields equations representing a charged wormhole that is
compatible with quantum field theory. \\
\phantom{a}\\
PAC numbers: 04.20.Jb, 04.20.Gz
\end{abstract}

\maketitle
\noindent

\section{Introduction}\noindent
Wormholes are handles or tunnels in the geometry of spacetime
connecting two distinct regions of our Universe or of completely
different universes.  The pioneer work of Morris and Thorne
\cite{MT88} has shown that, being solutions of the Einstein field
equations, macroscopic wormholes may be actual physical objects
that could even be traversed by humanoid travelers.  Unlike black
holes, which are also predictions of Einstein's theory, wormholes
can only be held open by the use of ``exotic" matter; such matter
violates the weak energy condition.

Because of the close connection between space and time,
general relativity is able to tolerate science-fiction type
phenomena such as wormholes and even time travel, as
exemplified by the G\"{o}del solution.  Quantum field theory, on
the other hand, is not so forgiving: it places severe
restrictions on the existence of traversable wormholes
\cite{FR05a, FR05b, Ford95, Ford96}.  In fact, according
to Ford and Roman \cite{Ford95, Ford96}, the wormholes
discussed in Ref. \cite{MT88} could not exist on a macroscopic
scale.  Interesting exceptions are the wormholes discussed
in Refs. \cite{MTY} and \cite{pK99}, but they are subject
to extreme fine-tuning.  This fine-tuning became an issue
in seeking compatibility with quantum field theory by a
suitable extension of the quantum inequalities
\cite{pK09, pK08}.  Given that exotic matter is rather
problematical, the idea behind the extension was to
strike a balance between reducing the size of the exotic
region and the concomitant fine-tuning of the metric
coefficients.  One can only be accomplished at the
expense of the other.

A particularly interesting generalization of the
Morris-Thorne wormhole can be obtained by the addition of an
electric charge, as proposed by Kim and Lee \cite{KL01, KK98}.
The resulting spacetime is a combination of a Morris-Thorne
spherically symmetric static wormhole and a
Reissner-N\"{o}rdstrom spacetime.

As in the case of black holes, wormholes with an electric
charge have been of interest for some time.  For example,
by adding an electric charge, Gonzales, Guzman, and
Sarbach \cite{GGS1} studied the possibility of stabilizing
a wormhole supported by a ghost scalar field, discussed in
their earlier papers \cite{GGS2, GGS3}.

Rotating and magnetized wormholes supported by phantom
scalar fields are discussed in Ref. \cite{tM}.  (A
ghost scalar field is often considered a simple example
of phantom energy, which is itself of interest in a
wormhole setting since it leads to a violation of the
weak energy condition.)

The aim of this paper is to show that a relatively small
modification of the metric describing a charged wormhole
suffices to satisfy an extended version of the Ford-Roman
inequality, thereby making such a wormhole compatible
with quantum field theory.  The modified model is also
a solution of the Einstein field equations.

\section{Traversable wormholes}\label{S:traversable}
\noindent
The spacetime geometry of a traversable wormhole can be
described by the metric

\begin{equation}\label{E:line1}
  ds^2=-e^{2\beta(r)}dt^2+e^{2\alpha(r)}dr^2+
        r^2(d\theta^2+\text{sin}^2\theta\,d\phi^2),
\end{equation}
where $\beta(r)\rightarrow 0$ and $\alpha(r)\rightarrow 0$ as
$r\rightarrow \infty$ and where $\alpha(r)$ has a continuous
derivative.  (We are using units in which $c=G=1$.)
The function $\beta=\beta(r)$ is called the \emph{redshift
function}, which must be everywhere finite to prevent an event
horizon.  The function $\alpha=\alpha(r)$ is related to the
\emph{shape function} $b=b(r)$:
\[
   e^{2\alpha(r)}=\frac{1}{1-b(r)/r}.
\]
So $b(r)=r(1-e^{-2\alpha(r)})$.  (Observe that $b'(r)$ is
continuous and that
$b(r)/r\rightarrow 0$ as $r\rightarrow \infty$.)  The minimum
radius $r=r_0$ is called the \emph{throat} of the wormhole,
where $b(r_0)=r_0$.  Also, $b'(r_0)\le 1$, referred to as
the \emph{flare-out condition} in Ref. \cite{MT88}.  It
follows that $\alpha$ has a vertical asymptote at $r=r_0$:
\[
  \lim_{r\to r_0+}\alpha(r)=+\infty.
\]

To hold a wormhole open, the weak energy condition (WEC) must be
violated.  The WEC states that the stress-energy tensor
$T_{\alpha\beta}$ must obey
\[
   T_{\alpha\beta}\mu^{\alpha}\mu^{\beta}\ge 0
\]
for all time-like vectors and, by continuity, all null
vectors.

\section{The quantum inequalities}\label{S:inequlities}
\noindent
To make this paper reasonably self-contained, we need a brief
discussion of the quantum inequalities due to Ford and Roman
\cite{Ford96}, slightly extended in \cite{pK09, pK08}.

In a series of papers, Ford and Roman (see Ref. \cite
{Ford96} and references therein) discuss a type of
constraint on the violation of the weak energy condition
by means of certain quantum inequalities which limit
the magnitude and time duration of negative energy.
These inequalities place severe restrictions on the
dimensions of Morris-Thorne wormholes.

One of these quantum inequalities, applied to different
situations, deals with an inertial Minkowski spacetime
without boundaries.  If $u^{\mu}$ is the observer's
four-velocity, that is, the tangent vector to a timelike
geodesic, then $\langle T_{\mu\nu}u^{\mu}u^{\nu}\rangle$
is the expectation value of the local energy density in
the observer's frame of
reference.  It is shown in Ref. \cite{Ford96} that
\begin{equation}\label{E:QI1}
   \frac{\tau_0}{\pi}\int^{\infty}_{-\infty}
   \frac{\langle T_{\mu\nu}u^{\mu}u^{\nu}\rangle d\tau}
    {\tau^2+\tau_0^2}\ge -\frac{3}{32\pi^2\tau_0^4}.
\end{equation}
Here $\tau$ is the observer's proper time and $\tau_0$ the
duration of the sampling time.  More precisely, the energy
density is sampled in a time interval of duration $\tau_0$
which is centered around an arbitrary point on the observer's
worldline so chosen that $\tau=0$ at this point.  [See Ref.
\cite{Ford96} for details.]

In a wormhole setting, a more convenient form is inequality
(\ref{E:QI3}) below, as we will see.  Applied to spherically
symmetric traversable wormholes in Ref. \cite{MT88}, it was
found that none were able to meet this condition.  As a
result, the throat sizes could only be slightly larger
than Planck length.  The inequality was subsequently
extended in Refs. \cite{pK09, pK08} to cover an entire
region around the throat.  It was then shown that it is
possible to strike a balance between the size of the
exotic region and the amount of fine-tuning required
to achieve this reduction.

Before discussing the extended quantum inequality, we
need to introduce the following length scales, modeled
after the length scales in Ref. \cite{Ford96}, which
were introduced in Ref. \cite {pK09}:
\begin{equation}\label{E:rsubm}
   r_m \equiv\text{min}\left[r,
   \left|\frac{b(r)}{b'(r)}\right|,
      \frac{1}{|\beta'(r)|},
   \left|\frac{\beta'(r)}{\beta''(r)}\right|\right].
\end{equation}
It is shown that if $R_{\text{max}}$ is the magnitude of
the maximum curvature, then $R_{\text{max}}\le 1/r^2_m$.
So the smallest radius of curvature $r_c$ is
\begin{equation*}
   r_c\approx \frac{1}{\sqrt{R_{\text{max}}}}\ge r_m.
\end{equation*}
Working on this scale, the spacetime is approximately
Minkowskian, so that inequality (\ref{E:QI1}) can be
applied with an appropriate $\tau_0$.

According to Ref. \cite{Ford96}, whenever the density $\rho$
is positive or zero, one should Lorentz transform to a
frame of a radially moving geodesic observer moving with
velocity $v$ relative to the static frame.  In this
``boosted frame" the density and maximum curvature are
denoted by $\rho'$ and $R'_{\text{max}}$, respectively.
We now have
\[
   r'_c\approx\frac{1}{\sqrt{R'_{\text{max}}}}\ge
      \frac{r_m}{\gamma},
\]
where $\gamma=(1-v^2)^{-1/2}$.  The suggested sampling
time is $\tau_0=fr_m/\gamma$, where $f$ is a scale factor
such that $f\ll 1$.  It is shown in Ref. \cite {pK09}
that in this boosted frame,
\begin{multline}\label{E:boost}
   T_{\hat{0}'\hat{0}'}=\rho'\\
   =\frac{\gamma^2}{8\pi r^2}\left[b'(r)
   -v^2\frac{b(r)}{r}+2v^2r\beta'(r)
    \left(1-\frac{b(r)}{r}\right)\right].
\end{multline}
In order for $\rho'$ to be negative, $v$ has to be
sufficiently large:
\begin{equation}\label{E:velocity}
  v^2>\frac{b'(r)}{b(r)/r-2r\beta'(r)
     \left(1-b(r)/r\right)}.
\end{equation}
Furthermore, inserting $l_p$,
\begin{multline}\label{E:QI2}
   \frac{r_m}{r}\le \\
    \left(\frac{1}{v^2b(r)/r-b'(r)-2v^2r\beta'(r)
    \left(1-b(r)/r\right)}\right)^{1/4}\\
       \times\frac{\sqrt{\gamma}}{f}
        \left(\frac{l_p}{r}\right)^{1/2}.
\end{multline}
At the throat, where $b(r_0)=r_0$, inequality (\ref{E:QI2})
reduces to Eq. (95) in Ref. \cite{Ford96}:
\begin{equation}\label{E:QI3}
  \frac{r_m}{r_0}\le\left(\frac{1}{v^2-b'(r_0)}\right)^{1/4}
       \frac{\sqrt{\gamma}}{f}\left(\frac{l_p}{r_0}\right)^{1/2}.
\end{equation}
Observe that inequality (\ref{E:QI3}) is trivially satisfied
if $b'(r_0)=1$, but not necessarily if $b'(r_0)<1$.  More
formally, inequality (\ref{E:QI3}) is satisfied whenever
$b'(r_0)-\epsilon<1$ for $\epsilon$ sufficiently small.  Since
$r_m$ includes $r_0$, the wormhole can be macroscopic.
Inequality (\ref{E:QI2}) will be applied in Sec.
\ref{S:feasibility}.

\emph{Remark 1:} To avoid division by zero in Eq. (\ref{E:QI3}),
we actually assume that $b'(r_0)$ is extremely close to 1
instead of exactly 1.  This also guarantees that the
flare-out condition is met at the throat \cite {MT88}:
\[
   \frac{b(r_0)-r_0b'(r_0)}{2[b(r_0)]^2}>0.
\]

Although retained here for convenience, the $v^2$ in
Eq. (\ref{E:QI2}) could actually be omitted \cite{pK09}.
The reason is that, according to Ref. \cite{FR05a}.
the boosted frame may be replaced by a static observer.

\section{The charged wormhole of Kim and Lee}\label{S:KL}
\noindent
To study a wormhole with a constant electric charge $Q$, it was
proposed by Kim and Lee \cite {KL01} that the Einstein field
equations take on the form
\[
   G^{(0)}_{\mu\nu}+G^{(1)}_{\mu\nu}=8\pi
      [T^{(0)}_{\mu\nu}+T^{(1)}_{\mu\nu}].
\]
In other words, the usual wormhole spacetime $G^{(0)}_{\mu\nu}
=8\pi T ^{(0)}_{\mu\nu}$ is to be modified by adding the matter
term $T^{(1)}_{\mu\nu}$ to the right side and the corresponding
back reaction $G^{(1)}_{\mu\nu}$ to the left side.  The proposed
metric is
\begin{multline}\label{E:line2}
  ds^2=-\left(1+\frac{Q^2}{r^2}\right)dt^2
   +\left(1-\frac{b(r)}{r}+\frac{Q^2}{r^2}\right)^{-1}dr^2\\
    +r^2(d\theta^2+\text{sin}^2\theta\,d\phi^2).
\end{multline}

Comparing metrics (\ref{E:line1}) and (\ref{E:line2}),
since $e^{2\beta(r)}=1+Q^2/r^2$, we have
\[
   \beta(r)=\frac{1}{2}\text{ln}\left(1+\frac{Q^2}{r^2}\right)
\]
and since $e^{2\alpha(r)}=1/[1-b(r)/r]$ in (\ref{E:line1}),
it follows that the effective shape function
$b_{\text{eff}}$ is
\begin{equation}\label{E:beff}
    b_{\text{eff}}(r)=b(r)-\frac{Q^2}{r}.
\end{equation}
Given the conversion factor $c^2/\sqrt{G}$, $Q^2$
is likely to be small in geometrized units.  Accordingly,
we will assume that $b(r)-Q^2/r>0$ in the vicinity of the
throat.

\section{The modified charged wormhole}\noindent
For reasons discussed later in this section, we are going to
propose the following modified metric for a charged wormhole:
\begin{multline}\label{E:line3}
  ds^2=-\left(1+R(r)+\frac{Q^2}{r^2}\right)dt^2\\
   +\left(1-\frac{b(r)}{r}+\frac{Q^2}{r^2}\right)^{-1}dr^2
    +r^2(d\theta^2+\text{sin}^2\theta\,d\phi^2).
\end{multline}
It is assumed that $R(r)\ge 0$, thereby avoiding an event
horizon, and that $R(r)$ has a continuous derivative.
As in metric (\ref{E:line2}), $b_{\text{eff}}(r)=b(r)-Q^2/r$.
Observe that the effective redshift function is
\[
  \beta(r)=\frac{1}{2}\text{ln}\left(1+R(r)+\frac{Q^2}{r^2}
        \right).
\]

In the discussion below, $\rho$ is the density, $\tau$ the radial
tension, and $p$ the transverse pressure.  Following Kim and Lee
\cite{KL01}, we assume that the matter terms are
\begin{equation}\label{E:WEC1}
   \rho^{(1)}=\tau^{(1)}=p^{(1)}=\frac{Q^2}{8\pi r^4}.
\end{equation}
The components of the Einstein tensor in the orthonormal
frame are
\begin{equation}\label{E:Einstein1}
   8\pi(\rho^{(0)}+\rho^{(1)})=\frac{b'}{r^2}+\frac{Q^2}{r^4},
\end{equation}
\begin{multline}\label{E:Einstein2}
  8\pi(\tau^{(0)}+\tau^{(1)})=\frac{b}{r^3}-\frac{Q^2}{r^4}\\
   -\frac{1-b/r+Q^2/r^2}{r(1+R(r)+Q^2/r^2)}
        \left[R'(r)-\frac{2Q^2}{r^3}\right],
\end{multline}
and
\begin{multline}\label{E:Einstein3}
   8\pi(p^{(0)}+p^{(1)})=\left(1-\frac{b}{r}+\frac{Q^2}{r^2}
                \right)\times\\
  \left[\beta''(r)-\frac{rb'-b+2Q^2/r}{2r(r-b+Q^2/r)}\beta'(r)
                   \right.\\
    \left.+[\beta'(r)]^2+\frac{\beta'(r)}{r}
    -\frac{rb'-b+2Q^2/r}{2r^2(r-b+Q^2/r)}\right].
\end{multline}
According to Ref. \cite{KL01}, employing the effective shape
function $b_{\text{eff}}(r)=b(r)-Q^2/r$ and assuming
$T^{\text{eff}}_{\mu\nu}=T^{(0)}_{\mu\nu}+T^{(1)}_{\mu\nu}$
(total matter) in the original Kim-Lee model, yields a
self-consistent solution of a system of equations similar
to that of the scalar field case discussed earlier in
Ref. \cite{KL01}.  The inclusion of the smooth function
$R(r)$ does not alter this conclusion.  So the metric
(\ref{E:line3}) may be viewed as a solution of the
Einstein field equations representing a wormhole with
an electric charge.

Returning to the WEC, if we use the radial outgoing null vector
$\mu^{\hat{\alpha}}=(1,1,0,0)$, then $T_{\hat{t}\hat{t}}+
T_{\hat{r}\hat{r}}=\rho-\tau\ge 0$.  For the above components,
since $\rho^{(1)}$ and $\tau^{(1)}$ drop out, we have
$\rho^{(0)}-\tau^{(0)}<0$ whenever the condition is violated.
(We will examine this violation shortly.)

Next, let us assume that $b=b(r)$ is a typical shape function in
the sense of Morris and Thorne \cite{MT88}: if the charge $Q$ is
zero, then the wormhole has a throat at $r=r*$, where $b(r*)=r*$.
For $r>r*$, we must have $b(r)<r$.  It follows that $b'(r*)\le 1$.
(See Fig. 1.)   The new shape function $b_{\text{eff}}(r)$
\begin{figure}[htbp]
\begin{center}
\includegraphics [clip=true, draft=false, bb= 0 0 305 190,
angle=0, width=4.5 in,
height=2.5 in, viewport=40 40 302 185]{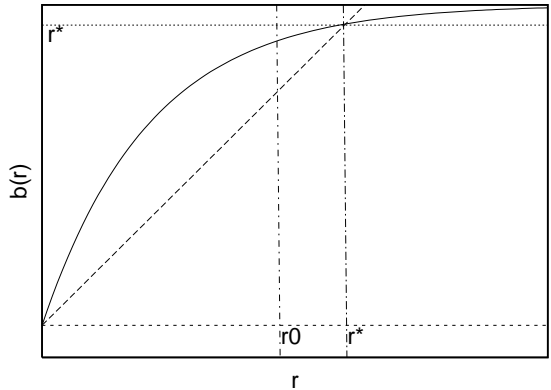}
\end{center}
\caption{\label{fig:figure1}The location of the throat
$r=r_0$.}
\end{figure}
[Eq.~(\ref{E:beff})] has analogous properties: in particular,
there is a throat at $r=r_0$, that is,
\begin{equation}\label{E:analogous}
   1-\frac{b(r_0)}{r_0}+\frac{Q^2}{r_0^2}=0.
\end{equation}
\emph{Remark 2:}  Eq. (\ref{E:analogous}) actually has two
roots,
\[
r_0=\frac{1}{2}\left(b(r_0)\pm\sqrt{[b(r_0)]^2-4Q^2}\right).
\]
In the special case $Q=0$, we get two possibilities,
$r_0=0$ and $r_0=b(r_0)$, showing that the smaller root is
meaningless.  We will therefore assume that there is only
one throat, corresponding to the larger root.

For $r>r_0$, $1-b(r)/r+Q^2/r^2>0$, or $b(r)/r-Q^2/r^2<1$.
Hence $b_{\text{eff}}(r)/r<1$ for $r>r_0$.  So we have, once
again, $b'_{\text{eff}}(r_0)\le 1$.  Finally, the profile
curve $z=z(r)$ is such that
\begin{equation}
   \frac{dz}{dr}=\pm\sqrt{
  \frac{b(r)/r-Q^2/r^2}{1-b(r)/r+Q^2/r^2}},
\end{equation}
showing that there is a vertical tangent at the throat $r=r_0$
in the usual embedding diagram.  As noted at the end of
Sec. \ref{S:KL}, the numerator is greater than zero.

Returning to $b=b(r)$, since $b(r)<r$ for $r>r*$, $b(r)>r$
for $r<r*$, and $b(r_0)-r_0=Q^2/r_0>0$ by
Eq.~(\ref{E:analogous}), it follows that $r_0<r*$.
(See Fig. 1.  If $b=b(r)$ is indeed a typical shape
function, then $b(r)>r$ for $r<r*$.)  Since we are using
geometrized units, $Q^2$ is likely to be small, so that
$r_0$ is not much less than $r*$.

The need for $b'_{\text{eff}}(r_0)\le 1$ referred to above
can be seen from the exoticity function in Ref. \cite{MT88}:
\begin{equation}
  \frac{b_{\text{eff}}(r_0)-r_0b'_{\text{eff}}(r_0)}
      {2[b_{\text{eff}}(r_0)]^2}>0.
\end{equation}
In other words, the flare-out condition is met whenever
\begin{equation}\label{E:flare1}
    \frac{b(r_0)}{r_0}-b'(r_0)>\frac{2Q^2}{r^2_0}
\end{equation}
at the throat.  This is consistent with the violation of the
WEC, $\rho^{(0)}-\tau^{(0)}<0$:
\begin{multline}
    8\pi(\rho^{(0)}-\tau^{(0)})=\frac{b'}{r^2}-\frac{b}{r^3}
     +\frac{2Q^2}{r^4}\\
     +\frac{1-b/r+Q^2/r^2}{r(1+R(r)+Q^2/r^2)}
     \left[R'(r)-\frac{2Q^2}{r^3}\right].
\end{multline}
At the throat,
\begin{equation}\label{E:flare2}
    \frac{r_0b'(r_0)-b(r_0)+2Q^2/r_0}{r_0^3}+0<0
\end{equation}
by inequality (\ref{E:flare1}).

\section{Feasibility}\label{S:feasibility}
\noindent
In order to study the feasibility of the charged wormhole, let us denote
the redshift function in the Kim and Lee wormhole by $\beta_1(r)$.
Thus
\begin{equation}\label{E:redshift1}
  \beta_1(r)=\frac{1}{2}\text{ln}\left(1+\frac{Q^2}{r^2}\right).
\end{equation}

Recall next that inequality (\ref{E:QI3}) is trivially satisfied if $b'=1$
at the throat.  For either of the charged wormholes, if $Q$ is chosen
properly, the condition can be met: from Eq.~(\ref{E:beff}),
\[
   b'_{\text{eff}}(r_0)=b'(r_0)+\frac{Q^2}{r_0^2};
\]
so for a proper choice of $Q$, $b'_{\text{eff}}(r_0)=1$,
even if $b'(r*)$ is less than 1.

To study the problem more closely, let us restate inequalities
(\ref{E:velocity}) and (\ref{E:QI2}) for $b_{\text{eff}}$:
\begin{equation}\label{E:velocity2}
    v^2>\frac{b'_{\text{eff}}(r)}
   {b_{\text{eff}}(r)/r-2r\beta'(r)(1-b_{\text{eff}}(r)/r)}
\end{equation}
and
\begin{multline}\label{E:QI4}
   \frac{r_m}{r}\le \\
    \left(\frac{1}{v^2b_{\text{eff}}(r)/r-b_{\text{eff}}'(r)-2v^2r\beta'(r)
    \left(1-b_{\text{eff}}(r)/r\right)}\right)^{1/4}\\
       \times\frac{\sqrt{\gamma}}{f}
        \left(\frac{l_p}{r}\right)^{1/2}.
\end{multline}
As before, at the throat, inequality (\ref{E:QI4}) is trivially satisfied
if $b_{\text{eff}}(r_0)=1.$

Problems arise when we move away from the throat.  It is shown in Ref.
\cite{pK09} that for any of the typical shape functions (which would
include $b_{\text{eff}}$), $b(r)/r-b'(r)>0$.  So for the wormhole in
Sec. \ref{S:KL}, the denominator on the right side of inequality
(\ref{E:QI4}) is no longer 0, since
\[
   \beta'_1(r)=-\frac{Q^2}{r(r^2+Q^2)}
\]
is negative.  Furthermore, with $Q$ fixed, $\beta(r)$ cannot be altered.
It is easy to demonstrate using specific shape functions that the quantum
inequalities cannot be met away from the throat.

To salvage the charged wormhole, some modification is
evidently needed.  With the Reissner-N\"{o}rdstrom
metric in mind, Eq.~(\ref{E:line3}) appears to be a natural
generalization, as long as $R(r)$ is not equal to $-b(r)/r$.
To distinguish this case from Eq.
(\ref{E:redshift1}), let us denote the modified redshift
function by $\beta_2(r)$:
\begin{equation}\label{E:redshift2}
    \beta_2(r)=\frac{1}{2}\text{ln}\left(1+R(r)
   +\frac{Q^2}{r^2}\right).
\end{equation}
The situation regarding the quantum inequalities is now
quite different:
\begin{equation}\label{E:beta2}
   \beta'_2(r)=\frac{1}{2}\frac{1}{1+R(r)+Q^2/r^2}
    \left[R'(r)-\frac{2Q^2}{r^3}\right],
\end{equation}
which is positive for a proper choice of $R(r)$.  Also,
$\beta'_2(r)$ is continuous if, and only if, $R'(r)$ is
continuous.  We proceed by first showing that
$\beta_2(r)$ (and hence $R(r)$) can be constructed or
adjusted to meet inequality (\ref{E:QI4}) away from the
throat.

Suppose we start at the throat and return to the
condition $b'_{\text{eff}}(r_0)=1$.  Then $v=1$ by
inequality (\ref{E:velocity2}), and inequality
(\ref{E:QI4}) is trivially satisfied.  Moving away
from the throat, the redshift function, now denoted
by $\beta(r)$, can be adjusted so that inequality
(\ref{E:QI4}) is still satisfied, i.e.,
\begin{equation}
   \frac{v^2b_{\text{eff}}(r)}{r}-b'_{\text{eff}}(r)-2rv^2
      \beta'_2(r)\left(1-\frac{b_{\text{eff}}(r)}{r}\right)
\end{equation}
is 0 or close to 0.  Assume that $\beta'(r)$ is
continuous.  (We also assume that the shape function
is fairly typical in the sense of having a gradually
decreasing slope, at least in the vicinity of the
throat.)

Substituting the newly determined $\beta(r)$ in
Eq. (\ref{E:beta2}), we get after rearranging,
\begin{equation}
   R'(r)-2\beta'(r)R(r)=2\beta'(r)+2\beta'(r)\frac{Q^2}{r^2}
      +\frac{2Q^2}{r^3}.
\end{equation}
The solution of the differential equation is
\begin{multline}\label{E:solution}
   R(r)=\\e^{2\beta(r)}\int^r_{r_0}e^{-2\beta(r')}
   \left(2\beta'(r')+2\beta'(r')\frac{Q^2}{(r')^2}
      +\frac{2Q^2}{(r')^3}\right)dr'.
\end{multline}
The continuity of $\beta'(r)$ is sufficient to guarantee
that $R(r)$ is a solution.  Using integration by parts,
the solution can be written
\begin{equation}
  R(r)=-1-\frac{Q^2}{r^2}+e^{2\beta(r)}e^{-2\beta(r_0)}
   \left(1+\frac{Q^2}{r_0^2}\right),
\end{equation}
showing that $R(r_0)=0$.  So it is in principle
possible to determine $R(r)$ from $\beta(r)$ such that
inequality (\ref{E:QI4}) is satisfied.  The resulting
wormhole will therefore become compatible with quantum 
field theory.

\section{Assigning various parameters-traversability}
\noindent
There are several parameters that come into play when
describing the wormhole geometry.  In particular,
$b_{\text{eff}}(r_0)=r_0$ and $b'_{\text{eff}}(r_0)=1$
lead to
\begin{equation}\label{E:condition1}
   b(r_0)-\frac{Q^2}{r_0}=r_0
\end{equation}
and
\begin{equation}\label{E:condition2}
   b'(r_0)+\frac{Q^2}{r_0^2}=1.
\end{equation}
If $b(r)$ and $Q$ are known, we can determine $r_0$.  It is
also possible to fix $r_0$ at some desired (macroscopic)
value and determine $b(r)$ and $Q$.  As a simple example,
suppose $b(r)$ has the form $b(r)=Ar^B$, $B<1$.  Then from
Eqs. (\ref{E:condition1}) and (\ref{E:condition2}), we
find that for nonzero $Q$,
\[
   A=\frac{2}{1+B}r_0^{1-B}\quad \text{and} \quad
    Q^2=\frac{1-B}{1+B}r_0^2.
\]
$B$ can be so chosen that $Q^2$ is relatively small, as
desired  in our geometrized units. This also confirms our
earlier assertion that $r_0$ cannot be much smaller than
$r*$ without making $Q^2$ unrealistically large.

As a check on the traversability by humanoid travelers (as
in Ref. \cite{MT88}), consider the proper distance $\ell(r)$
from the throat to a point away from the throat:
\[
   \ell(r)=\int^{r}_{r_0}\frac{dr'}
     {\sqrt{1-b'_{\text{eff}}(r')/r'}}.
\]
For $b(r)=Ar^B$ and $Q^2>0$, this distance is finite.  For example,
if $r_0=5$ m, and $Q^2=0.1$, then $\ell(6)\approx$ 140 m.
However, if $Q\rightarrow 0$, then $\ell(r)\rightarrow\infty$
for this particular shape function, so that the wormhole
would not be traversable.

\section{Conclusion}
\noindent
In this paper the charged wormhole described by the metric
\begin{multline*}
  ds^2=-\left(1+\frac{Q^2}{r^2}\right)dt^2
   +\left(1-\frac{b(r)}{r}+\frac{Q^2}{r^2}\right)^{-1}dr^2\\
    +r^2(d\theta^2+\text{sin}^2\theta\,d\phi^2)
\end{multline*}
due to Kim and Lee is extended to
\begin{multline*}
  ds^2=-\left(1+R(r)+\frac{Q^2}{r^2}\right)dt^2\\
   +\left(1-\frac{b(r)}{r}+\frac{Q^2}{r^2}\right)^{-1}dr^2
    +r^2(d\theta^2+\text{sin}^2\theta\,d\phi^2),
\end{multline*}
where $R(r)\ge 0$ has a continuous derivative at the throat, and
$R(r_0)=0$.  The main objective was to show that $R(r)$ can
be so chosen that the quantum inequality (\ref{E:QI4}) is satisfied
in the vicinity of the throat, thereby making the extended model 
compatible with quantum field theory.

It is also shown that the flare-out condition has been satisfied
and that an event horizon has been avoided.  Various combinations
of $b(r)$, $Q$, and $r=r_0$ are possible and may be chosen to make
the wormhole traversable by humanoid travelers.  The particular
model discussed shows that $Q$ may have to be nonzero for the
wormhole to be traversable.

\end{document}